\documentclass{article}
\usepackage{spconf,amsmath,graphicx}

\usepackage{color}
\usepackage{bm}
\usepackage{amssymb}
\usepackage[ruled,vlined]{algorithm2e}
\usepackage{multirow}
\usepackage{booktabs}
\usepackage{array}
\usepackage{lipsum}
\usepackage{enumerate}
\usepackage{stfloats}
\usepackage{subfigure}
\usepackage{cases}
\usepackage{diagbox}
\usepackage{cite}
\usepackage{balance}
\usepackage{halloweenmath}

\newcommand{\nid}{\noindent}

\title{Channel Estimation for Beyond Diagonal Reconfigurable \\ Intelligent Surfaces with Group-Connected Architectures}
\name{Hongyu~Li$^{\pumpkin}$, Yumeng~Zhang$^{\pumpkin}$, and Bruno~Clerckx$^{\pumpkin,\mathbat}$
}
\address{
$^{\pumpkin}$Department of Electrical and Electronic Engineering, Imperial College London, London, U.K.\\
$^{\mathbat}$Silicon Austria Labs (SAL), Graz, Austria\\
}

\begin{document}
%\ninept
%
\maketitle

\begin{abstract}
	We study channel estimation for a beyond diagonal reconfigurable intelligent surface (BD-RIS) aided multiple input single output system. 
    We first describe the channel estimation strategy based on the least square (LS) method, derive the mean square error (MSE) of the LS estimator, and formulate the BD-RIS design problem that minimizes the estimation MSE with unique constraints induced by group-connected architectures of BD-RIS. 
    Then, we propose an efficient BD-RIS design which theoretically guarantees to achieve the MSE lower bound. 
    Finally, we provide simulation results to verify the effectiveness of the proposed channel estimation scheme. 
\end{abstract}


\vspace{-0.3 cm}
\section{Introduction}
\label{sec:intro}
\vspace{-0.3 cm}

Beyond diagonal reconfigurable intelligent surface (BD-RIS) is a recently emerged technique, which goes beyond conventional RIS with diagonal phase shift matrices \cite{di2020smart,wu2019towards} and generates scattering matrices not limited to being diagonal, by introducing connections among RIS elements at the expense of increasing circuit complexity \cite{li2023reconfigurable}. 
Thanks to the flexible inter-element connections, BD-RIS has benefits in providing smarter wave manipulation and enlarging coverage \cite{li2023reconfigurable}. 

Existing works have been carried out for the modeling, beamforming design, and mode/architecture design of BD-RIS. 
The modeling of BD-RIS, involving the concept of group- and fully-connected architectures which are named according to the circuit topology of inter-element connections, is first proposed in \cite{shen2021}, followed by the discrete-value design and optimal beamforming design \cite{li2023reconfigurable}.
Inspired by the group/fully-connected architectures \cite{shen2021} and the concept of intelligent omni surface (IOS) or simultaneously transmitting and reflecting RIS (STAR-RIS) with enlarged coverage \cite{zhang2022intelligent,mu2021simultaneously}, BD-RIS with hybrid and multi-sector modes are proposed to achieve full-space coverage with enhanced performance \cite{li2023reconfigurable}. 
To find a better performance-complexity trade-off of BD-RIS, other architectures have also been investigated \cite{li2022reconfigurable,nerini2023beyond}. 
% Specifically, BD-RIS with dynamically group-connected architecture, where the grouping strategy is adaptive to the channel state information (CSI), is proposed in \cite{li2022dynamic} and has better performance than fixed group-connected BD-RIS \cite{li2022}.
Specifically, \cite{li2022reconfigurable} constructs BD-RIS with non-diagonal phase shift matrices relying on asymmetric circuit design to achieve higher channel gain than conventional RIS.
In addition, BD-RIS with tree- and forest-connected architectures are proposed in \cite{nerini2023beyond}, which are proved to achieve the performance upper bound with minimum circuit complexity.

\textit{Motivation:} The motivation of this work is twofold. 
1) The enhanced performance achieved by BD-RIS with different architectures highly depends on accurate channel state information (CSI), while none of the above-mentioned works study the channel estimation/acquisition of BD-RIS. 
2) Although the channel estimation protocol for conventional RIS scenarios \cite{yang2020intelligent, zheng2022survey,swindlehurst2022channel,you2020channel} still works in BD-RIS cases, the BD-RIS pattern for uplink training should be re-designed due to the different constraints on the scattering matrix, which yields different dimension and structure of the cascaded channel.

\textit{Contributions:} The contributions of this work are summarized as follows. 
\textit{First}, we propose a novel channel estimation scheme for a BD-RIS aided multiple input single output (MISO) system relying purely on the variation of BD-RIS matrix. 
\textit{Second}, we derive the lower bound of the mean square error (MSE) of the least square (LS) estimator and propose an efficient BD-RIS design to achieve the MSE lower bound. 
\textit{Third}, we present simulation results to verify the effectiveness and accuracy of the proposed channel estimation scheme.

\textit{Notations}:
% Boldface lower- and upper-case letters indicate column vectors and matrices, respectively.
% $\mathbb{E}\{\cdot\}$ represents the statistical expectation.
$(\cdot)^T$, $(\cdot)^*$, $(\cdot)^H$, and $(\cdot)^{-1}$ denote the transpose, conjugate, conjugate-transpose, and inversion operations, respectively.
$\otimes$ and $\odot$ denote the Kronecker product and Hadamard product, respectively. 
$\mathsf{blkdiag}(\cdot)$ is a block-diagonal matrix.
$\mathsf{vec}(\cdot)$, $\mathsf{rank}(\cdot)$, and $\mathsf{tr}(\cdot)$, respectively, are the vectorization, rank, and trace of a matrix.
$\overline{\mathsf{vec}}(\cdot)$ reshapes the vectorized matrix into the original matrix. 
$\mathsf{circshift}(\mathbf{a},N)$ rearranges vector $\mathbf{a}$ by moving the final $N$ entries to the first $N$ positions. 
$\mathsf{mod}(M,N)$ denotes $M$ modulo $N$. 

\vspace{-0.5 cm}
\section{System Model}
\label{sec:syst_mod}
\vspace{-0.3 cm}

\begin{figure}
    \centering
    \includegraphics[width=0.48\textwidth]{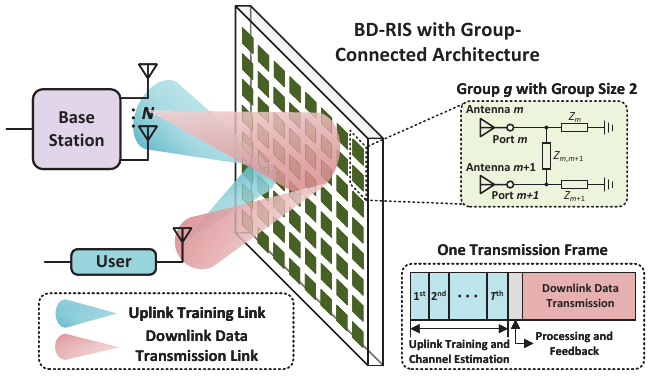}
    \caption{A paradigm of a BD-RIS assisted MISO system.}
    \label{fig:syst_mod}\vspace{-0.5 cm}
\end{figure}

We consider a narrowband system which consists of an $N$-antenna base station (BS), an $M$-antenna BD-RIS, and a single-antenna user, as illustrated in Fig. \ref{fig:syst_mod}. 
The $M$ antennas of the BD-RIS are connected to an $M$-port group-connected reconfigurable impedance network \cite{shen2021}, where the $M$ ports are uniformly divided into $G$ groups with each containing $\bar{M} = \frac{M}{G}$ ports connected to each other, $\bar{\mathcal{M}}=\{1,\ldots,\bar{M}\}$. 
Mathematically, the BD-RIS with group-connected architecture has a block-diagonal scattering matrix $\mathbf{\Phi} = \mathsf{blkdiag}(\mathbf{\Phi}_1,\ldots,\mathbf{\Phi}_G)\in\mathbb{C}^{M\times M}$ with each block $\mathbf{\Phi}_g\in\mathbb{C}^{\bar{M}\times\bar{M}}$ satisfying $\mathbf{\Phi}_g^H\mathbf{\Phi}_g = \mathbf{I}_{\bar{M}}$, $\forall g\in\mathcal{G} = \{1,\ldots,G\}$ \cite{shen2021}\footnote{When $\bar{M}=1$, the BD-RIS has a single-connected architecture and boils down to conventional RIS and the proposed channel estimation scheme is the same as the passive channel estimation for conventional RIS \cite{swindlehurst2022channel}.}. 
In this work, we assume the direct user-BS channel is blocked and focus purely on the estimation of the cascaded user-RIS-BS channel\footnote{When the direct BS-user channel exists, the direct channel can be effectively obtained by turning off the BD-RIS and using conventional channel estimation strategies.}.
Let $\mathbf{G}\in\mathbb{C}^{N\times M}$ and $\mathbf{h}\in\mathbb{C}^M$ denote the channel between the BD-RIS and BS, and between the user and BD-RIS, respectively. The user-RIS-BS channel $\mathbf{h}_\mathsf{u}\in\mathbb{C}^N$ is 
\begin{equation}
    \label{eq:channel}
    \mathbf{h}_\mathsf{u} = \mathbf{G}\mathbf{\Phi}\mathbf{h} = \sum_{g\in\mathcal{G}}\mathbf{G}_g\mathbf{\Phi}_g\mathbf{h}_g = \sum_{g\in\mathcal{G}}\underbrace{\mathbf{h}_g^T\otimes\mathbf{G}_g}_{=\mathbf{Q}_g\in\mathbb{C}^{N\times\bar{M}^2}}\mathsf{vec}(\mathbf{\Phi}_g),
\end{equation}
where $\mathbf{G}_g = [\mathbf{G}]_{:,(g-1)\bar{M}+1:g\bar{M}}\in\mathbb{C}^{N\times\bar{M}}$ and $\mathbf{h}_g = [\mathbf{h}]_{(g-1)\bar{M}+1:g\bar{M}}\in\mathbb{C}^{\bar{M}}$, $\forall g\in\mathcal{G}$. 
Expression (\ref{eq:channel}) indicates that the scattering matrix of BD-RIS with group-connected architecture is related to the cascaded channel $\mathbf{Q} = [\mathbf{Q}_1,\ldots,\mathbf{Q}_G]\in\mathbb{C}^{N\times\bar{M}^2G}$. This motivates us to directly estimate $\mathbf{Q}$ instead of separate channels and perform the beamforming design with the knowledge of $\mathbf{Q}$ for data transmission, which results in the following protocol \cite{yang2020intelligent,swindlehurst2022channel,you2020channel} with each transmission frame divided into three phases as illustrated in Fig. \ref{fig:syst_mod}.  

\textit{Phase 1:} The BS estimates $\mathbf{Q}$ by uplink training, where the pilots are consecutively transmitted from the user and reflected by the BD-RIS with varied scattering matrix $\mathbf{\Phi}$ during the training period. In this phase, the difference compared to conventional RIS comes from the design of the varied RIS pattern due to the different constraint of the scattering matrix, whose details will be given in Section \ref{sec:CE}.

\textit{Phase 2:} The BS optimizes the transmit precoder and BD-RIS matrix based on the estimated cascaded channel, and feeds back the results to the user and BD-RIS. 
This phase is also different from conventional RIS cases due to the more general unitary constraint of the BD-RIS. 
Specifically, with determined cascaded channel $\mathbf{Q}$, the received power maximization problem to jointly design the transmit precoder $\mathbf{w}\in\mathbb{C}^N$ and BD-RIS matrix $\mathbf{\Phi}'$ for the downlink MISO is
\begin{equation}
    \label{eq:opt_prob}
    % \begin{align}
    \max_{\mathbf{\Phi}_g'^H\mathbf{\Phi}_g' = \mathbf{I}_{\bar{M}},\|\mathbf{w}\|^2\le P} ~~ \Big|\sum_{g\in\mathcal{G}}\mathsf{vec}^T(\mathbf{\Phi}_g'^T)\mathbf{Q}_g^T\mathbf{w}\Big|^2,
    % \label{eq:ris_constraint}
    % \mathrm{s.t.} ~~~&\mathbf{\Phi}_g'^H\mathbf{\Phi}_g' = \mathbf{I}_{\bar{M}}, \forall g\in\mathcal{G},~\|\mathbf{w}\|^2 \le P,
    % \end{align}
\end{equation}
where $P$ is the transmit power.
The main difficulty in solving problem (\ref{eq:opt_prob}) lies in the non-convex constraint of BD-RIS. The solution for problem (\ref{eq:opt_prob}) will be further investigated in the journal extension.

\textit{Phase 3:} The BS performs the downlink data transmission based on the optimized precoder and BD-RIS matrix. 

\textit{Remark 1.}
The feasibility of this protocol is supported by the following aspects. 1) The communication occurs with time division duplex (TDD) and the reciprocity between the uplink and downlink channel exists. This indicates that the downlink channel can be expressed using the uplink cascaded channel $\mathbf{Q}$, i.e., $\mathbf{h}_\mathsf{d} = \mathbf{h}'\mathbf{\Phi}'\mathbf{G}' = \sum_{g\in\mathcal{G}}\mathbf{h}_g'\mathbf{\Phi}_g'\mathbf{G}_g' = \sum_{g\in\mathcal{G}}(\mathbf{h}_g'\otimes\mathbf{G}_g'^T\mathsf{vec}(\mathbf{\Phi}_g'^T))^T \overset{\text{(a)}}{=} \sum_{g\in\mathcal{G}}(\mathbf{h}_g^T\otimes\mathbf{G}_g\mathsf{vec}(\mathbf{\Phi}_g'^T))^T = \sum_{g\in\mathcal{G}}\mathsf{vec}^T(\mathbf{\Phi}_g'^T)\mathbf{Q}_g^T$, where (a) holds due to the reciprocity between $\mathbf{h}$ and $\mathbf{h}'$, and $\mathbf{G}$ and $\mathbf{G}'$, i.e., $\mathbf{h}' = \mathbf{h}^T$ and $\mathbf{G}' = \mathbf{G}^T$.
2) The cascaded channel $\mathbf{Q}$ remains approximately constant within one transmission frame. 3) The knowledge of the cascaded channel is sufficient for beamforming design as formulated in Phase 2.

\vspace{-0.3 cm}
\section{Channel Estimation for BD-RIS}
\label{sec:CE}
\vspace{-0.3 cm}

In this section, we focus on Phase 1 and describe the proposed channel estimation strategy based on the LS method, formulate the BD-RIS design problem to minimize the MSE of the LS estimator, and provide the solution. 

\vspace{-0.3 cm}
\subsection{LS Based Channel Estimation}
\vspace{-0.2 cm}

The channel estimation strategy is described as follows. Assuming the user sends pilot symbol $x_{t}\in\mathbb{C}$, $|x_t|=1$ at time slot $t$, $\forall t\in\mathcal{T} = \{1,\ldots,T\}$, the signal received at the BS is 
\begin{equation}\label{eq:received_signal_uplink}
    \begin{aligned}
        \mathbf{y}_t &= \sqrt{P_\mathrm{u}}\sum_{g\in\mathcal{G}}\mathbf{Q}_g\mathsf{vec}(\mathbf{\Phi}_{t,g})x_t + \mathbf{n}_t\\
        &= \sqrt{P_\mathrm{u}}\mathbf{Q}\widehat{\bm{\phi}}_t + \mathbf{n}_t, \forall t\in\mathcal{T},
    \end{aligned}
\end{equation}
where $P_\mathrm{u}$ denotes the transmit power at the user, $\mathbf{\Phi}_{t,g}\in\mathbb{C}^{\bar{M}\times\bar{M}}$ denotes the $g$-th block of the BD-RIS matrix at time slot $t$, and $\mathbf{n}_t \in\mathbb{C}^N$ denotes the noise with $\mathbf{n}_t\sim\mathcal{CN}(\mathbf{0},\widehat{\sigma}^2\mathbf{I}_N)$, $\forall t\in\mathcal{T}$.
The vector $\widehat{\bm{\phi}}_t$ is defined as $\widehat{\bm{\phi}}_t = [\mathsf{vec}^T(\mathbf{\Phi}_{t,1}),\ldots,\mathsf{vec}^T(\mathbf{\Phi}_{t,G})]^Tx_{t}\in\mathbb{C}^{G\bar{M}^2}$, $\forall t\in\mathcal{T}$.
To uniquely estimate the cascaded channel $\mathbf{Q}$ in (\ref{eq:received_signal_uplink}), $T$ pilot symbols should be transmitted from the user. Without loss of optimality, we assume $x_t = 1$, $\forall t\in\mathcal{T}$ and combine the data from such pilots together, which yields 
\begin{equation}
    \begin{aligned}
        \mathbf{Y} = [\mathbf{y}_1,\ldots,\mathbf{y}_T]
        = \sqrt{P_\mathrm{u}}\mathbf{Q}\underbrace{[\widehat{\bm{\phi}}_1,\ldots,\widehat{\bm{\phi}}_T]}_{=\widehat{\mathbf{\Phi}}\in\mathbb{C}^{G\bar{M}^2\times T}} + \underbrace{[\mathbf{n}_1,\ldots,\mathbf{n}_T]}_{=\mathbf{N}\in\mathbb{C}^{N\times T}}.
    \end{aligned}
\end{equation}
The simplest way to estimate $\mathbf{Q}$ is to use the LS method, yielding the LS estimator of $\mathbf{Q}$ as 
\begin{equation}
    \widehat{\mathbf{Q}} = (\sqrt{P_\mathrm{u}})^{-1}\mathbf{Y}\widehat{\mathbf{\Phi}}^\dagger = (\sqrt{P_\mathrm{u}})^{-1}\mathbf{Y}\widehat{\mathbf{\Phi}}^H(\widehat{\mathbf{\Phi}} \widehat{\mathbf{\Phi}}^H)^{-1},
\end{equation}
with $T\ge G\bar{M}^2$ to guarantee the recovery of $\widehat{\mathbf{Q}}$. 
Then the MSE of the LS estimator is 
\begin{equation}
    \label{eq:mse_ls_estimate}
    \mathrm{e}_{\widehat{\mathbf{Q}}} = \mathbb{E}\{\|\widehat{\mathbf{Q}} - \mathbf{Q}\|_F^2\} = \frac{N\widehat{\sigma}^2}{P_\mathrm{u}}\mathsf{tr}((\widehat{\mathbf{\Phi}}\widehat{\mathbf{\Phi}}^H)^{-1}),
\end{equation} 
which implies that the MSE of channel estimation depends purely on the value of matrix $\widehat{\mathbf{\Phi}}$ such that a proper design of $\widehat{\mathbf{\Phi}}$ is required. As such, we formulate the following MSE minimization problem 
\begin{subequations}\label{eq:prob_MSE}
    \begin{align}
    \label{eq:obj_MSE}
    \min_{\widehat{\mathbf{\Phi}}}~ &\mathsf{tr}((\widehat{\mathbf{\Phi}}\widehat{\mathbf{\Phi}}^H)^{-1})\\
    \label{eq:unitary_constraint}
    \text{s.t.}~~ &\mathbf{\Phi}_{t,g}^H\mathbf{\Phi}_{t,g} = \mathbf{I}_{\bar{M}}, \forall t\in\mathcal{T},\forall g\in\mathcal{G},\\
    \label{eq:rank}
    &\mathsf{rank}(\widehat{\mathbf{\Phi}}) = G\bar{M}^2,
    \end{align}
\end{subequations}
where we set $T$ as $T^\mathrm{min}=G\bar{M}^2$ to minimize the overhead without corrupting the MSE performance, which yields $\widehat{\mathbf{\Phi}}$ a full-rank square matrix. 
Problem (\ref{eq:prob_MSE}) is difficult to solve due to the inverse operation in the objective and the non-convex constraints of group-connected BD-RIS. In the following subsection, we will first simplify the objective function and then propose an efficient approach for optimal $\widehat{\mathbf{\Phi}}$. 

\vspace{-0.3 cm}
\subsection{Solution to Problem (\ref{eq:prob_MSE})}
\vspace{-0.2 cm}

We start by deriving the lower bound of the objective (\ref{eq:obj_MSE}) with constraint (\ref{eq:unitary_constraint}) based on the following lemma. 

\textit{Lemma 1.}
The objective (\ref{eq:obj_MSE}) has the following lower bound $\mathsf{tr}((\widehat{\mathbf{\Phi}}\widehat{\mathbf{\Phi}}^H)^{-1}) \ge \bar{M}$, where the equality achieves when $\widehat{\mathbf{\Phi}}\widehat{\mathbf{\Phi}}^H = \widehat{\mathbf{\Phi}}^H\widehat{\mathbf{\Phi}}= M\mathbf{I}_{G\bar{M}^2}$.

\textit{Proof.}
The objective (\ref{eq:obj_MSE}) has the following lower bound 
\begin{equation}
	\label{eq:lower_bound}
	\mathsf{tr}((\widehat{\mathbf{\Phi}}\widehat{\mathbf{\Phi}}^H)^{-1}) = \sum_{i=1}^{G\bar{M}^2}[(\widehat{\mathbf{\Phi}}\widehat{\mathbf{\Phi}}^H)^{-1}]_{i,i} \overset{\text{(a)}}{\ge}\sum_{i=1}^{G\bar{M}^2}\frac{1}{[\widehat{\mathbf{\Phi}}\widehat{\mathbf{\Phi}}^H]_{i,i}},
\end{equation}
where the equality of (a) can be attained when $\widehat{\mathbf{\Phi}}\widehat{\mathbf{\Phi}}^H$ is a diagonal matrix \cite{kay1993fundamentals}. Additionally, we have 
\begin{equation}\label{eq:HM_AM}
	\begin{aligned}
	\sum_{i=1}^{G\bar{M}^2}\frac{1}{[\widehat{\mathbf{\Phi}}\widehat{\mathbf{\Phi}}^H]_{i,i}} &\overset{\text{(b)}}{\ge} \frac{G^2\bar{M}^4}{\sum_{i=1}^{G\bar{M}^2}[\widehat{\mathbf{\Phi}}\widehat{\mathbf{\Phi}}^H]_{i,i}} \\
	&= \frac{G^2\bar{M}^4}{\mathsf{tr}(\widehat{\mathbf{\Phi}}\widehat{\mathbf{\Phi}}^H)} = \frac{G^2\bar{M}^4}{\mathsf{tr}(\widehat{\mathbf{\Phi}}^H\widehat{\mathbf{\Phi}})}
	\overset{\text{(c)}}{=}\bar{M},
	\end{aligned}
\end{equation}
where (b) holds due to the relationship between the harmonic mean and the arithmetic mean with equality achieved when $[\widehat{\mathbf{\Phi}}\widehat{\mathbf{\Phi}}^H]_{1,1}=\ldots=[\widehat{\mathbf{\Phi}}\widehat{\mathbf{\Phi}}^H]_{G\bar{M}^2,G\bar{M}^2}$; (c) holds due to the constraint (\ref{eq:unitary_constraint}) which yields $[\widehat{\mathbf{\Phi}}^H\widehat{\mathbf{\Phi}}]_{1,1} = \ldots = [\widehat{\mathbf{\Phi}}^H\widehat{\mathbf{\Phi}}]_{G\bar{M}^2,G\bar{M}^2} = M$. Combining (\ref{eq:lower_bound}) and (\ref{eq:HM_AM}), we can achieve the lower bound $\mathsf{tr}((\widehat{\mathbf{\Phi}}\widehat{\mathbf{\Phi}}^H)^{-1}) = \bar{M}$ with $\widehat{\mathbf{\Phi}}\widehat{\mathbf{\Phi}}^H = \widehat{\mathbf{\Phi}}^H\widehat{\mathbf{\Phi}}=M\mathbf{I}_{G\bar{M}^2}$, which completes the proof. 
\hfill  $\square$

With Lemma 1, we can transform problem (\ref{eq:prob_MSE}) into the following feasibility-check problem 
\begin{subequations}\label{eq:feasibility_check}
    \begin{align}
        \text{find} ~~&\widehat{\mathbf{\Phi}}\in\mathbb{C}^{G\bar{M}^2\times G\bar{M}^2}\\
        \label{eq:fc_cons1}
        \text{s.t.}~~ &\widehat{\mathbf{\Phi}}\widehat{\mathbf{\Phi}}^H = M\mathbf{I}_{G\bar{M}^2},
        \text{(\ref{eq:unitary_constraint})}.
    \end{align}
\end{subequations}
\nid Given the large dimension of matrix $\widehat{\mathbf{\Phi}}$, it is difficult to directly find such a matrix to simultaneously satisfy the two constraints in problem (\ref{eq:feasibility_check}). To reduce the dimension of designed variables and simplify the design of matrix $\widehat{\mathbf{\Phi}}$, we give the following lemma to further transform problem (\ref{eq:feasibility_check}).

\textit{Lemma 2.} 
The feasible matrix $\widehat{\mathbf{\Phi}}$ from problem (\ref{eq:feasibility_check}) can be constructed as $\widehat{\mathbf{\Phi}} = \mathbf{X}\otimes\bar{\mathbf{\Phi}}$, where $\mathbf{X}\in\mathbb{C}^{G\times G}$ is obtained by solving the following problem:

\vspace{-0.5 cm} 
\begin{subequations}\label{eq:find_X}
    \begin{align}       
        \text{find} ~~&\mathbf{X}\in\mathbb{C}^{G\times G}\\
        \label{eq:fx_cons1}
        \text{s.t.} ~~&\mathbf{X}\mathbf{X}^H = G\mathbf{I}_{G},\\
        \label{eq:fx_cons2}
        &|[\mathbf{X}]_{g,g'}| = 1, \forall g,g'\in\mathcal{G}.
    \end{align}
\end{subequations} 
\vspace{-0.5 cm}

\nid $\bar{\mathbf{\Phi}}\in\mathbb{C}^{\bar{M}^2\times\bar{M}^2}$ is obtained by solving the following problem:

\vspace{-0.5 cm} 
\begin{subequations}\label{eq:find_Phy}
    \begin{align}       
        \text{find} ~~&\bar{\mathbf{\Phi}}\in\mathbb{C}^{\bar{M}^2\times\bar{M}^2}\\
        \label{eq:fc1_cons1}
        \text{s.t.} ~~&\bar{\mathbf{\Phi}}\bar{\mathbf{\Phi}}^H = \bar{M}\mathbf{I}_{\bar{M}^2},\\
        \label{eq:fc1_cons2}
        &\overline{\mathsf{vec}}^H([\bar{\mathbf{\Phi}}]_{:,m})\overline{\mathsf{vec}}([\bar{\mathbf{\Phi}}]_{:,m}) = \mathbf{I}_{\bar{M}}, \forall m\in\bar{\bar{\mathcal{M}}},
    \end{align}
\end{subequations}
\vspace{-0.5 cm} 

\nid where $\bar{\bar{\mathcal{M}}} = \{1,\ldots,\bar{M}^2\}$.

\textit{Proof.} 
With $\mathbf{X}$ satisfying (\ref{eq:fx_cons1}), (\ref{eq:fx_cons2}) and $\bar{\mathbf{\Phi}}$ satisfying (\ref{eq:fc1_cons1}), (\ref{eq:fc1_cons2}), we have $\widehat{\mathbf{\Phi}} = \mathbf{X}\otimes\bar{\mathbf{\Phi}}$ such that
$\widehat{\mathbf{\Phi}}\widehat{\mathbf{\Phi}}^H = (\mathbf{X}\otimes\bar{\mathbf{\Phi}})(\mathbf{X}^H\otimes\bar{\mathbf{\Phi}}^H) = (\mathbf{X}\mathbf{X}^H)\otimes(\bar{\mathbf{\Phi}}\bar{\mathbf{\Phi}}^H) = (G\mathbf{I}_G)\otimes(\bar{M}\mathbf{I}_{\bar{M}^2}) = M\mathbf{I}_{G\bar{M}^2}$,
which aligns with (\ref{eq:fc_cons1}).
In addition, each block of $\widehat{\mathbf{\Phi}}$, i.e., $\widehat{\mathbf{\Phi}}_{g,g'} = [\widehat{\mathbf{\Phi}}]_{(g-1)\bar{M}^2+1:g\bar{M}^2,(g'-1)\bar{M}^2+1:g'\bar{M}^2} = [\mathbf{X}]_{g,g'}\bar{\mathbf{\Phi}}$, $\forall g,g'\in\mathcal{G}$, is constructed by columns $[\widehat{\mathbf{\Phi}}_{g,g'}]_{:,m} = \mathsf{vec}(\mathbf{\Phi}_{(g'-1)\bar{M}^2+m,g}) = \mathsf{vec}(\mathbf{\Phi}_{t,g})$ satisfying (\ref{eq:unitary_constraint}), that is, 
$\mathbf{\Phi}_{t,g}^H\mathbf{\Phi}_{t,g} =\overline{\mathsf{vec}}^H([\widehat{\mathbf{\Phi}}_{g,g'}]_{:,m})\overline{\mathsf{vec}}([\widehat{\mathbf{\Phi}}_{g,g'}]_{:,m})
= |[\mathbf{X}]_{g,g'}|^2\times\overline{\mathsf{vec}}^H([\bar{\mathbf{\Phi}}]_{:,m}) \overline{\mathsf{vec}}([\bar{\mathbf{\Phi}}]_{:,m})=\mathbf{I}_{\bar{M}}$. 
The proof is completed.
\hfill $\square$

With Lemma 2, we decouple problem (\ref{eq:feasibility_check}) into two reduced-dimensional problems (\ref{eq:find_X}) and (\ref{eq:find_Phy}). 
The feasible solution to problem (\ref{eq:find_X}) can be easily obtained by using a $G\times G$ discrete Fourier transform (DFT) matrix $\mathbf{F}_G$ or a Hadamard matrix $\mathbf{D}_G$, i.e., $\mathbf{X} = \mathbf{F}_G$ or $\mathbf{X} = \mathbf{D}_G$. However, the solution to problem (\ref{eq:find_Phy}) is not that straightforward since we need to find a matrix $\bar{\mathbf{\Phi}}$ such that each column constructs a unitary matrix, i.e., constraint (\ref{eq:fc1_cons2}) and that different columns are orthogonal to each other, i.e., constraint (\ref{eq:fc1_cons1}). 
Therefore, the matrix $\bar{\mathbf{\Phi}}$ should include two-dimensional orthogonality, which motivates us to construct $\bar{\mathbf{\Phi}}$ with two orthogonal bases. 
This brings the following theorem. 

\textit{Theorem 1.}
The matrix $\bar{\mathbf{\Phi}}$ satisfying (\ref{eq:fc1_cons1}), (\ref{eq:fc1_cons2}) can be constructed such that each column, i.e., $[\bar{\mathbf{\Phi}}]_{:,(m-1)\bar{M}+n} = \bar{\bm{\phi}}_{m,n}$, $\forall m,n\in\bar{\mathcal{M}}$, has the following structure:
\begin{equation}
    \begin{aligned}
    \bar{\bm{\phi}}_{m,n} = \mathsf{circshift}(\mathsf{vec}(\mathbf{Z}_1),(n-1)\bar{M})\odot([\mathbf{Z}_2]_{:,m}\otimes\mathbf{1}_{\bar{M}}),
    \end{aligned}
\end{equation}
where $\mathbf{Z}_1\in\mathbb{C}^{\bar{M}\times\bar{M}}$ is a scaled unitary matrix, i.e., $\mathbf{Z}_1^H\mathbf{Z}_1=\alpha_1\mathbf{I}_{\bar{M}}$, $\mathbf{Z}_2\in\mathbb{C}^{\bar{M}\times\bar{M}}$ is a scaled unitary matrix whose entries have identical modulus, i.e., $\mathbf{Z}_2^H\mathbf{Z}_2=\alpha_2\mathbf{I}_{\bar{M}}$, $|[\mathbf{Z}_2]_{m,n}|=\sqrt{\frac{\alpha_2}{\bar{M}}}$, $\forall m,n\in\bar{\mathcal{M}}$, and $\alpha_1\alpha_2 = \bar{M}$.

\textit{Proof.}
We start by proving $\bar{\bm{\phi}}_{m,n}$ satisfies (\ref{eq:fc1_cons2}). To this end, we block $\bar{\bm{\phi}}_{m,n} = [\bar{\bm{\phi}}_{m,n,1}^H,\bar{\bm{\phi}}_{m,n,2}^H,\ldots,\bar{\bm{\phi}}_{m,n,\bar{M}}^H]^H$ with $\bar{\bm{\phi}}_{m,n,i} = [\bar{\bm{\phi}}_{m,n}]_{(i-1)\bar{M}+1:i\bar{M}} = [\mathbf{Z}_2]_{i,m}[\mathbf{Z}_1]_{:,\mathsf{mod}(i-n,\bar{M})+1}$ and calculate
% \begin{equation}
%     \begin{aligned}
%         \non
        $\bar{\bm{\phi}}_{m,n,i}^H\bar{\bm{\phi}}_{m,n,i'}
        =z_{2,i,i',m}z_{1,i.i',n}$, $z_{1,i.i',n}=[\mathbf{Z}_1]_{:,\mathsf{mod}(i-n,\bar{M})+1}^H[\mathbf{Z}_1]_{:,\mathsf{mod}(i-n,\bar{M})+1}$, $z_{2,i,i',m}=[\mathbf{Z}_2]_{i,m}^*\times[\mathbf{Z}_2]_{i',m}$.
%     \end{aligned}
% \end{equation}
This yields the following two conditions:
1) when $i=i'$, we have $z_{1,i,i,n} = \alpha_1$ and $z_{2,i,i,m} = \frac{\alpha_2}{\bar{M}}$ such that $\bar{\bm{\phi}}_{m,n,i}^H\times\bar{\bm{\phi}}_{m,n,i'}=1$;
2) when $i\ne i'$, we have $z_{1,i,i',n} = 0$ such that $\bar{\bm{\phi}}_{m,n,i}^H\bar{\bm{\phi}}_{m,n,i'} = 0$. Therefore, (\ref{eq:fc1_cons2}) is guaranteed.

We next prove $\bar{\mathbf{\Phi}}$ constructed by $\bar{\bm{\phi}}_{m,n}$ satisfies (\ref{eq:fc1_cons1}). To this end, we calculate 
% \begin{equation}
%     \begin{aligned}
%         \non
        $\bar{\bm{\phi}}_{m,n}^H\bar{\bm{\phi}}_{m',n'} 
        = \sum_{i=1}^{\bar{M}}\bar{\bm{\phi}}_{m,n,i}^H\bar{\bm{\phi}}_{m',n',i}
        = \sum_{i=1}^{\bar{M}}z_{2',m,m',i}z_{1',n,n',i}$, $z_{2',m,m',i}=[\mathbf{Z}_{2}]_{i,m}^*[\mathbf{Z}_{2}]_{i,m'}$, and $z_{1',n,n',i}=[\mathbf{Z}_1]_{:,\mathsf{mod}(i-n,\bar{M})+1}^H[\mathbf{Z}_1]_{:,\mathsf{mod}(i-n',\bar{M})+1}$.
%     \end{aligned}
% \end{equation}
This yields the following three conditions:
1) when $m=m'$ and $n=n'$, we have $z_{1',n,n,i} = \alpha_1$ and $z_{2',m,m,i} = \frac{\alpha_2}{\bar{M}}$ such that $\bar{\bm{\phi}}_{m,n}^H\bar{\bm{\phi}}_{m,n} = \bar{M}$; 
2) when $n\ne n'$, we have $z_{1',n,n',i} =0$ such that $\bar{\bm{\phi}}_{m,n}^H\bar{\bm{\phi}}_{m',n'} = 0$;
3) when $m\ne m'$ and $n=n'$, we have $z_{1',n,n,i} = \alpha_1$ such that $\bar{\bm{\phi}}_{m,n}^H\bar{\bm{\phi}}_{m',n'} = \alpha_1\sum_{i=1}^{\bar{M}}z_{2',m,m',i} = \alpha_1[\mathbf{Z}_2]_{:,m}^H[\mathbf{Z}_2]_{:,m'} = 0$. Therefore, (\ref{eq:fc1_cons1}) is guaranteed. The proof is completed.
\hfill $\square$

According to Theorem 1, we can simply choose $\mathbf{Z}_1 = \mathbf{F}_{\bar{M}}$ or $\mathbf{Z}_1 = \mathbf{D}_{\bar{M}}$ and $\mathbf{Z}_2 = \frac{1}{\sqrt{\bar{M}}}\mathbf{F}_{\bar{M}}$ or $\mathbf{Z}_2 = \frac{1}{\sqrt{\bar{M}}}\mathbf{D}_{\bar{M}}$ to construct $\bar{\mathbf{\Phi}}$, and further construct $\widehat{\mathbf{\Phi}} = \mathbf{X}\otimes\bar{\mathbf{\Phi}}$ with $\mathbf{X} = \mathbf{F}_G$ or $\mathbf{X}=\mathbf{D}_G$. In this way, we achieve the MSE lower bound of the LS estimation, i.e., 
$\mathrm{e}_{\widehat{\mathbf{Q}}}^\mathrm{min} = \frac{N\widehat{\sigma}^2}{P_\mathrm{u}}\bar{M}$ and avoid the time-consuming inversion operation with $\widehat{\mathbf{\Phi}}^\dag = \frac{1}{M}\widehat{\mathbf{\Phi}}^H$.

\vspace{-0.3 cm}
\subsection{Discussion}
\vspace{-0.2 cm}

The minimum overhead to estimate $\mathbf{Q}$ is given by $T^\mathrm{min} = G\bar{M}^2$, which depends both on the number of groups $G$ and on the group size $\bar{M}$ of the group-connected BD-RIS. More specifically, $T^\mathrm{min}$ grows faster with $\bar{M}$ than with $G$. Meanwhile, the estimation error $\mathrm{e}^\mathrm{min}_{\widehat{\mathbf{Q}}}$ increases with $\bar{M}$. These two observations indicate that the proposed estimation method is more appealing with a small group size $\bar{M}$.  

\vspace{-0.3 cm}
\section{Performance Evaluation}
\label{sec:simulation}
\vspace{-0.3 cm}

In this section, we perform simulation results to verify the effectiveness of the proposed channel estimation scheme. 
The simulation parameters are set as follows. 
The BS is equipped with $N=4$ antennas. The BD-RIS is equipped with $M = 32$ antennas. 
Both BS-RIS and RIS-user channels are assumed to have Rayleigh fading\footnote{In this work, we consider i.i.d. Rayleigh fading for simplicity to verify the accuracy of the theoretical derivations, while the proposed scheme also works when spatial correlation exists.} accounting for the small-scale fading and distance-dependent model accounting for the large-scale fading, i.e., $\zeta_{o}=\zeta_0(\frac{d_i}{d_0})^{-\varepsilon}$, $\forall o\in\{\mathrm{BI,IU}\}$. $\zeta_0 = -30$ dB is the signal attenuation at reference distance $d_0 = 1$ m. BS-RIS and RIS-user distances are respectively set as $d_\mathrm{BI} = 50$ m and $d_\mathrm{IU} = 10$ m. The pathloss component is set as $\varepsilon = 2.2$. The noise power is set as $\widehat{\sigma}^2 = -100$ dBm.

\begin{figure}
    \centering
    \includegraphics[width=0.48\textwidth]{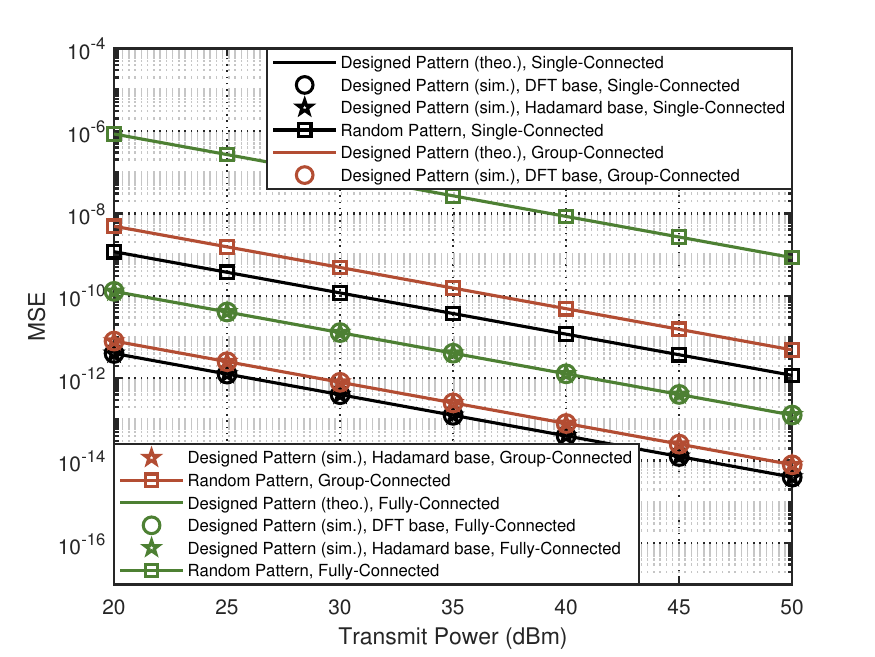}
    \vspace{-0.5 cm}
    \caption{MSE versus $P$ for BD-RIS with single-, group-, and fully-connected architectures (The group size for group- and fully-connected architectures are $\bar{M} =2$ and $\bar{M}=M$).}
    \label{fig:MSE_P}\vspace{-0.5 cm}
\end{figure}

We plot the MSE\footnote{In this work, we plot the MSE performance instead of normalized MSE (NMSE) to compare the proposed scheme with the theoretical lower bound.} versus transmit power for conventional (single-connected) RIS and BD-RIS with group- and fully-connected ($\bar{M}=M$) architectures in Fig. \ref{fig:MSE_P}, from which we have the following observations. 
\textit{First}, the matrix $\widehat{\mathbf{\Phi}}$ with DFT/Hadamard bases achieves exactly the same MSE performance as the theoretical lower bound, which verifies Lemma 2 and Theorem 1. 
\textit{Second}, the MSE increases with the group size $\bar{M}$ of group-connected BD-RIS, which also verifies Lemma 1. 
\textit{Third}, the MSE decreases with transmit power, which indicates that larger transmit power leads to smaller channel estimation error. 
\textit{Fourth}, the matrix $\widehat{\mathbf{\Phi}}$ with DFT/Hadamard bases achieves better MSE performance than that constructed by random unitary matrices, which demonstrates the effectiveness of the proposed BD-RIS design for channel estimation.

\vspace{-0.4 cm}
\section{Conclusion}
\label{sc:Conclusion}
\vspace{-0.3 cm}

We propose a novel channel estimation scheme for a BD-RIS aided MISO system. Specifically, the BD-RIS has a group-connected architecture, which, mathematically, generates unique constraints and complicates the BD-RIS design and thus the channel estimation. To tackle this difficulty, we first derive the MSE lower bound of the LS estimator. Then, we propose an efficient BD-RIS design to achieve the MSE lower bound. 
Finally, simulation results demonstrate the superiority of the proposed design. 
In the near future, it is worth investigating more efficient channel estimation schemes for BD-RIS, with smaller estimation error and lower overhead. 

\addtolength{\voffset}{-0.01 in}

\vfill\pagebreak
	

\balance
\bibliographystyle{IEEEbib}
\bibliography{refs}

\begin{thebibliography}{10}

\bibitem{di2020smart}
Marco Di~Renzo, Alessio Zappone, Merouane Debbah, Mohamed-Slim Alouini, Chau
  Yuen, Julien De~Rosny, and Sergei Tretyakov,
\newblock ``Smart radio environments empowered by reconfigurable intelligent
  surfaces: How it works, state of research, and the road ahead,''
\newblock {\em IEEE Journal on Selected Areas in Communications}, vol. 38, no.
  11, pp. 2450--2525, 2020.

\bibitem{wu2019towards}
Qingqing Wu and Rui Zhang,
\newblock ``Towards smart and reconfigurable environment: Intelligent
  reflecting surface aided wireless network,''
\newblock {\em IEEE communications magazine}, vol. 58, no. 1, pp. 106--112,
  2019.

\bibitem{li2023reconfigurable}
Hongyu Li, Shanpu Shen, Matteo Nerini, and Bruno Clerckx,
\newblock ``Reconfigurable intelligent surfaces 2.0: Beyond diagonal phase
  shift matrices,''
\newblock {\em arXiv preprint arXiv:2301.03288}, 2023.

\bibitem{shen2021}
Shanpu Shen, Bruno Clerckx, and Ross Murch,
\newblock ``Modeling and architecture design of reconfigurable intelligent
  surfaces using scattering parameter network analysis,''
\newblock {\em IEEE Transactions on Wireless Communications}, vol. 21, no. 2,
  pp. 1229--1243, 2021.

\bibitem{zhang2022intelligent}
Hongliang Zhang, Shuhao Zeng, Boya Di, Yunhua Tan, Marco Di~Renzo, M{\'e}rouane
  Debbah, Zhu Han, H~Vincent Poor, and Lingyang Song,
\newblock ``Intelligent omni-surfaces for full-dimensional wireless
  communications: Principles, technology, and implementation,''
\newblock {\em IEEE Communications Magazine}, vol. 60, no. 2, pp. 39--45, 2022.

\bibitem{mu2021simultaneously}
Xidong Mu, Yuanwei Liu, Li~Guo, Jiaru Lin, and Robert Schober,
\newblock ``Simultaneously transmitting and reflecting {(STAR) RIS} aided
  wireless communications,''
\newblock {\em IEEE Transactions on Wireless Communications}, vol. 21, no. 5,
  pp. 3083--3098, 2021.

\bibitem{li2022reconfigurable}
Qingchao Li, Mohammed El-Hajjar, Ibrahim~A Hemadeh, Arman Shojaeifard, Alain
  Mourad, Bruno Clerckx, and Lajos Hanzo,
\newblock ``Reconfigurable intelligent surfaces relying on non-diagonal phase
  shift matrices,''
\newblock {\em IEEE Transactions on Vehicular Technology}, vol. 71, no. 6, pp.
  6367--6383, 2022.

\bibitem{nerini2023beyond}
Matteo Nerini, Shanpu Shen, Hongyu Li, and Bruno Clerckx,
\newblock ``Beyond diagonal reconfigurable intelligent surfaces utilizing graph
  theory: Modeling, architecture design, and optimization,''
\newblock {\em arXiv preprint arXiv:2305.05013}, 2023.

\bibitem{yang2020intelligent}
Yifei Yang, Beixiong Zheng, Shuowen Zhang, and Rui Zhang,
\newblock ``Intelligent reflecting surface meets {OFDM}: Protocol design and
  rate maximization,''
\newblock {\em IEEE Transactions on Communications}, vol. 68, no. 7, pp.
  4522--4535, 2020.

\bibitem{zheng2022survey}
Beixiong Zheng, Changsheng You, Weidong Mei, and Rui Zhang,
\newblock ``A survey on channel estimation and practical passive beamforming
  design for intelligent reflecting surface aided wireless communications,''
\newblock {\em IEEE Communications Surveys \& Tutorials}, vol. 24, no. 2, pp.
  1035--1071, 2022.

\bibitem{swindlehurst2022channel}
A~Lee Swindlehurst, Gui Zhou, Rang Liu, Cunhua Pan, and Ming Li,
\newblock ``Channel estimation with reconfigurable intelligent surfaces--{A}
  general framework,''
\newblock {\em Proceedings of the IEEE}, 2022.

\bibitem{you2020channel}
Changsheng You, Beixiong Zheng, and Rui Zhang,
\newblock ``Channel estimation and passive beamforming for intelligent
  reflecting surface: Discrete phase shift and progressive refinement,''
\newblock {\em IEEE Journal on Selected Areas in Communications}, vol. 38, no.
  11, pp. 2604--2620, 2020.

\bibitem{kay1993fundamentals}
Steven~M Kay,
\newblock {\em Fundamentals of statistical signal processing: estimation
  theory},
\newblock Prentice-Hall, Inc., 1993.

\end{thebibliography}
	
\end{document}